\pdfoutput=1
\documentclass{PoS_oldRef}
\let\ifpdf\relax
\usepackage{upgreek}

\title{Transient Astrophysics with the Square Kilometre Array}

\ShortTitle{SKA Transients}

\author{\speaker{Rob Fender}\\
	Oxford Astrophysics\\
        E-mail: \email{rob.fender@astro.ox.ac.uk}}

\author{Adam Stewart\\
	Oxford Astrophysics\\
	E-mail: \email{adam.stewart@astro.ox.ac.uk}}

\author{Jean-Pierre Macquart\\
        ICRAR/Curtin\\
        E-mail: \email{J.Macquart@curtin.edu.au}}

\author{Immacolata Donnarumma\\
	IAPS/INAF Rome\\
	E-mail: \email{immacolata.donnarumma@iaps.inaf.it}}

\author{Tara Murphy\\
	University of Sydney\\
	E-mail: \email{tara@physics.usyd.edu.au}}

\author{Adam Deller\\
	ASTRON\\
	E-mail: \email{deller@astron.nl}}

\author{Zsolt Paragi\\
	Joint Institute for VLBI in Europe\\
	E-mail: \email{zparagi@jive.nl}}

\author{Shami Chatterjee\\
	Cornell\\
	E-mail: \email{shami@astro.cornell.edu}}

\abstract{
This chapter provides an overview of the possibilities for transient and variable-source astrophysics with the Square Kilometre Array. While subsequent chapters focus on the astrophysics of individual events, we focus on the broader picture, and how to maximise the science coming from the telescope. The SKA as currently designed will be a fantastic and ground-breaking facility for radio transient studies, but the scientifc yield will be dramatically increased by the addition of (i) near-real-time commensal searches of data streams for events, and (ii) on occasion, rapid robotic response to Target-of-Opprtunity style triggers.
}

\FullConference{
Advancing Astrophysics with the Square Kilometre Array\\
June 8-13, 2014\\
Giardini Naxos, Italy}

\newcommand{\skipthis}[1]{}

\begin{document}

\section{Introduction}

Radio transients are both the sites and signatures of the most extreme phenomena in our Universe: e.g. exploding stars, compact object mergers, black holes and ultra-relativistic flows. 
They also have the potential to act as probes of the intervening medium on all scales, up to and including cosmological distances. As such they are invaluable probes for subjects as diverse as stellar evolution, relativistic astrophysics and cosmology.

The majority of such transients can be broadly divided into:

\begin{itemize}
\item{
{\bf Incoherent synchrotron events}, which are associated with all explosive kinetic feedback and particle acceleration in the Universe, from relativistic jets to supernova explosions. Such emission is limited, in a steady state, to a brightness temperature of $T_{\mathrm{B}} \leq 10^{12}$K, and therefore the most luminous events (observable at cosmological distances) vary rather slowly (days to years).

 This timescale is often longer than that of a typical observation, and variability is usually determined by comparing images made at different epochs. As noted below, synchrotron flares can generally be used as a measure of the kinetic feedback associated with an event, and as such allow us to understand the flow of energy in the most extreme astrophysical environments.}

\item{
{\bf Coherent bursts}, which can achieve much higher brightness temperature (up to at least $10^{30}$K) and hence can be of very short durations. Such events can be intrinsically much shorter than the timescales imposed by scattering and dispersion in the intervening interstellar and intergalactic medium, and so can be used as unique probes of these environments. These events are generally observed in `pulsar mode' beamformed data.}
\end{itemize}

The correspondence between emission mechansim and observing modes is not exact, however, and very fast imaging (on millisecond timescales) and strong scattering can blur the boundaries between the two. Furthermore, while the above constitute the majority of radio transients, strong variability can also be observed associated with thermal emission (e.g. novae), gravitational lensing and scattering, all of which can be picked up by the revised approaches outlined here.  In all cases, transient science benefits greatly from monitoring the sky frequently.  As such, commensal observing alongside other projects is a powerful strategy.

\subsection{Examples of important science associated with radio variables}

There are many examples of unexepected and/or extreme astrophysical phenomena being revealed or better understood via observations of radio variability. These include:

\begin{itemize}
\item{Radio bursts from Jupiter (Burke \& Franklin 1955).}
\item{The discovery of neutron stars via their pulsed radio emission (Hewish et al. 1968).}
\item{Discovery of apparent superluminal motion in relativistic jets in variable extragalactic (Cohen et al. 1971) and Galactic (`microquasars', Mirabel \& Rodriguez 1994) sources (black holes).}
\item{Extreme scattering events caused by small scale structures in the ISM (Fiedler et al. 1987).}
\item{The powerful, beamed, jet-like nature of gamma-ray bursts, and their probable association with unusual supernovae (Kulkarni et al. 1998).}
\item{Elucidation of the relation between accretion states and feedback in accreting black holes (Fender et al. 2004).}
\item{Discovery of Fast Radio Bursts (Lorimer et al. 2007).}
\item{Highly relativistic jet-like flows associated with the tidal disruption and accretion of a star (Zauderer et al. 2011).}
\end{itemize}

Wilkinson (2015, this volume) discusses other cases of serendipitous discoveries made in the radio band.

\begin{figure}[h]
\includegraphics[width=.95\textwidth, angle=0]{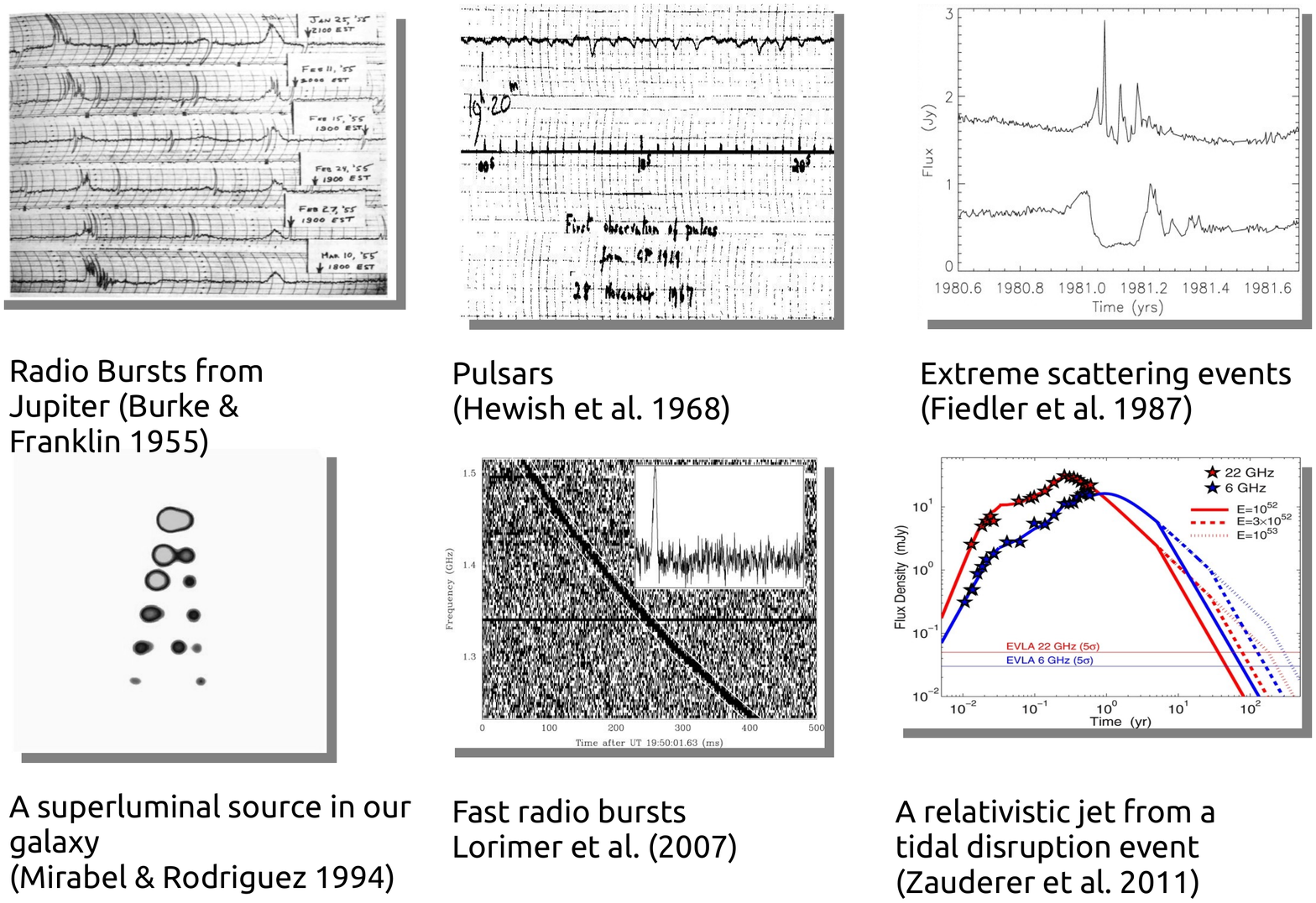}
\caption{Some, but by no means all, of the major astrophysical discoveries associated with variable radio emission. Major discoveries have been made entirely serendipitously (e.g. pulsars), by searching parameter space (e.g. Fast Radio Bursts) and by rapidly following-up events discovered at other wavelengths (e.g. Microquasars, jetted Tidal Disruption Events).}
\label{greatest}
\end{figure}

Of course transient studies are given high priority in other wavelength bands. X-ray astronomy has a rich history of discovering variable and transient X-ray sources as well as reporting them in real time to the global community (e.g. RXTE All-Sky Monitor, Levine et al. 1996; {\em Swift}, Gehrels \& Cannizzo 2015). Transient science is undergoing a golden age in the optical band (e.g. the Palomar Transients Factory, Law et al. 2009) and is key science for the LSST in the 2020s (Abell et al. 2009).

The key advantages and complementarity of radio emission compared to other wavelengths include:

\begin{itemize}

\item{{\bf Measuring kinetic feedback, which often originates in relativistically moving ejecta:} radio emission uniquely traces the feedback of kinetic energy, via relativistic jets or more isotropic explosions. Such explosions result in the acceleration of particles and compression of magnetic fields, resulting in synchrotron emission from which the minimum injected energy can be directly estimated.}

\item{{\bf Probing the properties of the intervening ionised media:} radio waves are scattered, dispersed and have their plane of polarisation rotated as they propagate through various media along the line of sight. These effects deliver more and complementary information about the properties of intergalactic and interstellar space than do the absorption and reddening observed at higher frequencies.}

\item{{\bf Precise localisation across very wide fields of view:} the beauty of radio interferometers is the ability to use small dishes or even dipoles to survey very large areas of sky whilst simultaneously having high angular resolution (the downside of this being the very large computational resource required to image huge numbers of pixels). The wide fields of view ($\rightarrow$ high survey speed) are essential for fast and efficient searches, the excellent potential localisation important for selecting candidates at other wavelengths in crowded fields (e.g. globular clusters or galaxies).}

\end{itemize}

\section{Incoherent (image plane) transients}

\begin{figure}[h]
\includegraphics[width=.99\textwidth, angle=0]{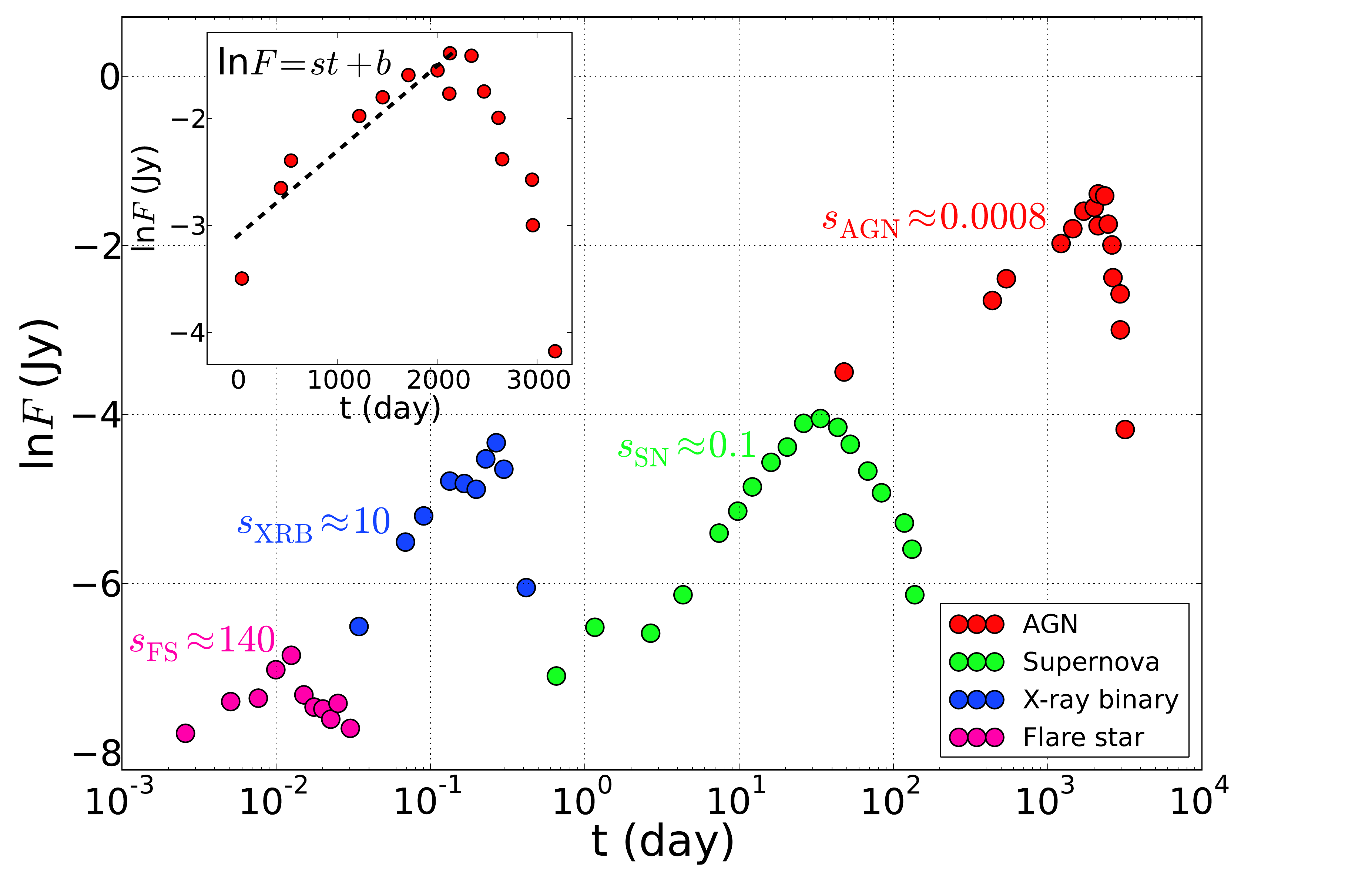}
\caption{Incoherent synchrotron flares on a range of timescales from different objects. 
The fitted constant $s$, corresponds to $\tau^{-1}$, where $\tau$ is the exponential rise time of the events.
The evolution is often adequately described  (in terms of the reliably extracted astrophysics) by the simple expanding-blob model of van der Laan (1966).
From Pietka, Fender \& Keane (2015).}
\label{Gosia2}
\end{figure}

Essentially all explosive events in astrophysics are associated with incoherent synchrotron emission, resulting from ejections at velocities in excess of the local sound speed which compress ambient magnetic fields and accelerate particles. These events range from relatively low-luminosity flares from stars, to the most powerful events in the Universe, associated with gamma-ray bursts and relativistic jets from super-massive black holes in active galactic nuclei. 
Fig \ref{Gosia2} presents a set of synchrotron flares associated with different types of object and different timescales: more luminous objects evolve more slowly. This is expected because incoherent synchrotron emission in a steady state is limited to a (rest-frame) brightness temperature of $T_{\mathrm{B}} \sim 10^{12}$K, and so the more luminous sources must be physically larger and hence vary more slowly.

\begin{figure}[h]
\includegraphics[width=.99\textwidth, angle=0]{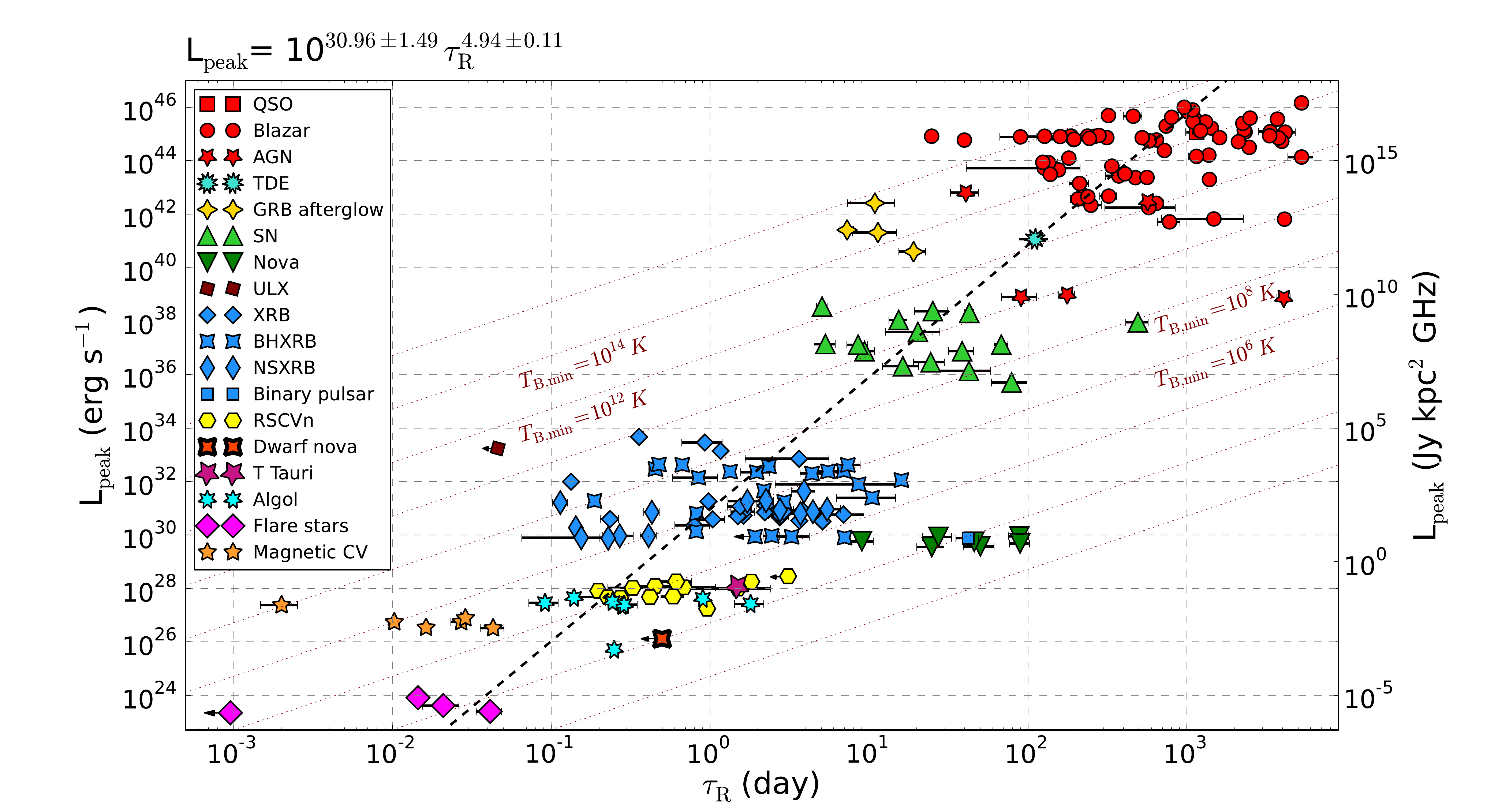}
\caption{Comparison of radio luminosity with rise time for a (biased) sample of sources for which the distance is known. The peak radio luminosity is a function of of the flare rise timescale. The diagonal (red) lines correspond to a fixed minimum brightness temperature (which would imply $L \propto \tau^2$), and so we see that the more luminous objects also have higher (minimum) brightness temperatures.
This result is biased, at least partially, by the beaming of some of the most luminous sources. From Pietka, Fender \& Keane (2015), which includes a more comprehensive discussion of these biases.
}
\label{Gosia3}
\end{figure}

The best-fit relation between peak radio luminosity and flare timescale for a (biased) sample of sources with known distances (and hence radio luminosities) is presented in Fig \ref{Gosia3} and is approximately of the form

\[
L_{\rm radio} \propto \tau^5
\]

where $\tau$ is the exponential rise timescale (Pietka, Fender \& Keane 2015). The results are very similar for decays.

Such synchrotron events typically have rich multiwavelength counterparts, and can be associated with explosions and accretion events which are visible at optical, X-ray and gamma-ray wavelengths. Combining these data sets can provide unique insights with, as noted earlier, the radio providing both a key estimate of kinetic feedback and, in some cases, more precise localisation and resolved ejecta. There is also usually a delay in the sense that the highest-energy emission escapes while the synchrotron source is still optically thick at most radio wavelengths and hence the radio emission peaks much later. For many sources the timescale of the trigger event will be much shorter than the timescale for the source to evolve to an optically thin phase at radio wavelengths, and so the delay between event and radio peak will be approximately the same as $\tau$ (See Figs 2, 3). There may be exceptions to this however, such as relatively slowly-evolving internal shocks in the jets of variable velocities.

For more details on the important science and exciting prospects for these kinds of events, we refer the interested reader to Corbel et al. (2015, this volume) and Yu et al. (2015, this volume) for radio emission from accreting binaries and related systems, Donnarumma et al. (2015, this volume) for jetted tidal disruption events, Burlon et al. (2015, this volume) for gamma-ray bursts, Perez-Torres et al. (2015, this volume) and Wang et al. (2015, this volume) for supernovae. Variable radio emission (both intrinsic and propagation-induced) associated with AGN is discussed in Bignall et al. (2015, this volume), and radio emission from thermal sources such as novae in O'Brien et al. (2015, this volume).

\subsection{Coherent (beamformed) transients}

\begin{figure}[h]
\includegraphics[width=.99\textwidth, angle=0]{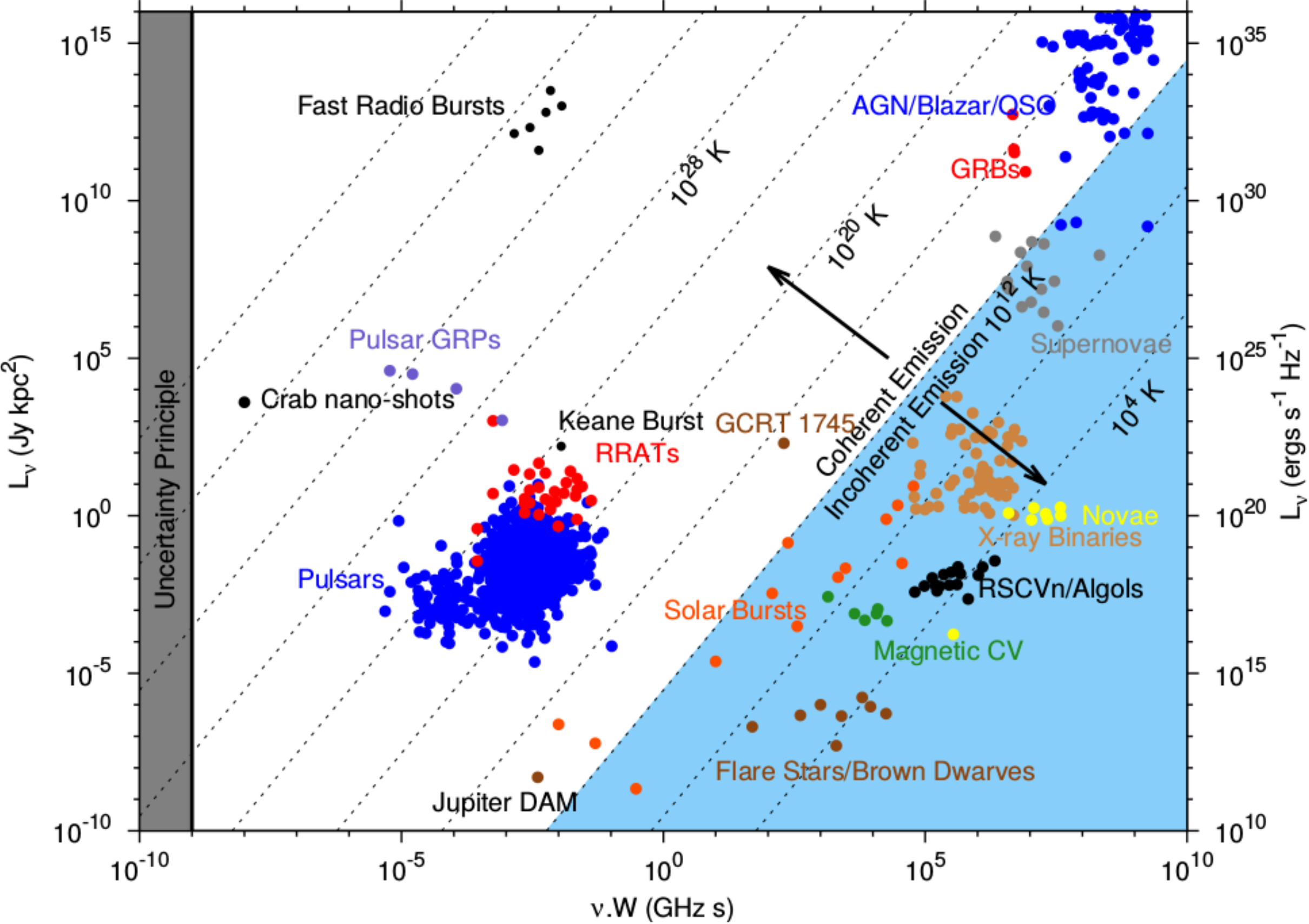}
\caption{Transients parameter space expanded to include coherent sources. From Pietka, Fender \& Keane (2015), following a long line of similar plots (e.g. Cordes, Lazio \& McLaughlin 2004). }
\label{Gosia5}
\end{figure}

Coherent radio emission can have a much higher brightness temperature than incoherent synchrotron emission (see Fig \ref{Gosia5}), and is exemplified by pulsars which typically have $T_{\mathrm{B}} \geq 10^{20}$K (see Kramer \& Stappers, 2015; this volume, and references therein for the exciting prospects for pulsar science with the SKA); other physical mechanisms include cyclotron masers. Coherent flares are not, typically, associated with energy release on the scale of the most luminous synchrotron events, and are more commonly used as precise probes of the physical conditions at the emission site as well as propagation effects in the intervening ISM/IGM.

Fig \ref{Gosia5} expands the parameter space explored in Fig \ref{Gosia3} to include also coherent events, including Solar and Jovian events from within our own solar system, the shortest timescale events observed from pulsars, and Fast Radio Bursts. 

\section{Rates}

The rate of transient events at a given depth, cadence and observing frequency is not well measured in the radio band (in fact it is not well reported in most bands of the electromagnetic spectrum). The overwhelming majority of radio sources observed by the SKA will be distant galaxies, a large fraction of which will be AGN. Low-level variability of these AGN, both intrinsic and via scintillation (see also Bignall et al. 2015, this volume) will result in at least 1\% of radio sources being variables at the tens of percent level (e.g. Deller \& Middelberg 2014).

However, relatively low-level AGN activity, while interesting, is not our main target. In searching for, and following up, radio transients, we wish both to study populations we know to exist (for example GRBs) and to search for new and unexpected populations. In terms of the known populations, we choose to use, as an indicator of the likely success rate of the SKA, two topical and exciting classes of object, the Fast Radio Bursts (FRBs) and the (jetted) Tidal Disruption Events (TDEs). We should be clear that we are focussing on these two classes of object simply as examples: transient science is too diverse to devote space to a dedicated discussion to all of the very interesting astrophysics under study (e.g. X-ray binaries, GRBs, stellar flares).

\subsection{Fast Radio Bursts}

Fast Radio Bursts (FRBs) are very short ($\leq 15$ ms), apparently coherent, bursts which show
large excess dispersion when compared to that expected for the line of sight within our galaxy.
First discovered by Lorimer et al. (2007), if the excess dispersion is associated with the IGM, they
could lie at cosmological distances (the current sample may be probing to redshift $z \sim 1$), be the
tracers of new astrophysical phenomena and acts as probes of the missing baryons (see discussion
in Macquart et al. 2015, this volume).

Initial estimates placed the FRB rate at $1.0^{+0.6}_{-0.5} \times 10^4\,$events\,sky$^{-1}$day$^{-1}$ for bursts above the flux density threshold of the Parkes radio telescope, whose $10 \sigma$ sensitivity reaches 460\,mJy in 1\,ms (Thornton et al.\,2013).   However, subsequent FRB detections in the High Time Resolution Universe survey (HTRU; Keith et al. 2010) reveal a significant disparity in the detection rate between events detected above and below a Galactic latitude of $30^\circ$, with the event rate being $\approx 4$ times higher at high latitudes (Petroff et al.\,2014).

Uncertainties in the FRB flux density distribution and spectrum render the estimation of event detection rates for other telescopes uncertain by several orders of magnitude (but see Lorimer et al. 2013 \& Trott et al. 2013).  However, a simple estimate of the event detection rate at observing frequencies comparable to those at which FRB detections are made (i.e. 1.2-1.7\,GHz, as probed by SKA1-MID) is possible based on reasonable estimates of the slope of the differential flux density distribution (i.e. the log $N$-log $S$ distribution).  The cumulative event rate scales as $\Omega S_0^{-3/2}$ for events that are distributed homogeneously throughout a Universe in which spacetime curvature is neglected.  

SKA1-MID should be a potent FRB detection machine.  
With an SEFD of 1.7\,Jy and a bandwidth of 720\,MHz over the 900-1670\,MHz observing band, SKA1-MID reaches a $10\sigma$ sensitivity of $S_0=14$\,mJy in 1\,ms.  This represents a factor of 33 sensitivity advantage over Parkes.  For an SKA1-MID comprised of $15\,$m diameter dishes, it possesses a field of view ($\Omega$) advantage of a factor of 1.4 over the Parkes multibeam receiver.  Thus, if SKA1-MID is able to conduct a tied-array beam search for FRBs over the entire field of view out to the half-power point of the primary beam, this represents a factor $\Omega S_0^{-3/2} \approx 260$ advantage over the Parkes HTRU survey.

However, recent critical examination of the cause of the FRB event rate dependence on Galactic latitude suggests that such simple estimates are in error (Macquart \& Johnston, submitted).  Diffractive interstellar scintillation of FRB radiation can enhance common, weaker events at high Galactic latitudes to a point where they are detectable, thus enhancing the apparent even rate under certain circumstances. If this effect is the correct explanation of the FRB latitude event rate disparity, it also implies that the FRB differential flux density distribution is much steeper than expected, with the cumulative event rate scaling between $\Omega \, S_0^{-2.4}$ and $\Omega \, S_0^{-2.8}$.  

Applying the foregoing considerations implies SKA1-MID should detect FRBs at a rate $(6-25) \times 10^3$ higher than the Parkes rate.

\subsection{Tidal Disruption Events}

Tidal disruption events (TDEs) are accretion events onto a central galactic black hole (e.g. dormant AGN) which result from the tidal disruption and accretion of (approximately half of) a star which passes close to the black hole. The events are faster and of higher amplitude than typical AGN activity, and may be our closest analogues to the instability-induced outbursts in stellar mass black hole binaries, which had led to a new understanding of the connection between relativistic accretion and ejection. 
Zauderer et al. (2011) reported the detection of a radio flare from such an event, likely to have been associated with the production of a relativistic jet (for more discussion see Donnarumma et al. 2015, this volume).
SKA / SKA1-MID will have great potential to discover and follow up large numbers of such events.
This is achievable by means of a commensal search in all SKA observations, or by carrying out a near real-time transient search in a high cadence dedicated survey. The latter is preferred to detect TDEs early on in order to allow for the necessary high energy follow-ups aimed at confirming the TDE origin of the transient event. In the following, we discuss both real-time and commensal search for the discovery. 

In the first case, we define a trade-off between sky coverage and sensitivity, favoring short integration times in order to allow for detection at higher energies and then enable the identification of the transient at higher redshift (see Donnarumma et al. 2015). We assume a multiple cadence half-sky survey, comparable to the shallow surveys proposed by the `continuum' working group (Prandoni et al. 2015, this volume).
By assuming an half sky coverage with a 2-day cadence at a $5-\sigma$ flux limit of  $90 ~\upmu \rm Jy$, we estimate a number between $\sim 300$ and $\sim 800$  yr$^{-1}$ (a bulk Lorentz factor $\Gamma=2$ is assumed). This is because the 1.4 GHz light curve of the prototype event, Sw J1644 increases with time at least up to 600 days (Zauderer et al. 2013), therefore a multiple cadence survey with longer integration times ($\sim 8$ days) needed to achieve the same flux limit of our strategy should not result in a significant loss of radio detections. However,  radio triggers at late times could be problematic for follow-up observations, because sources would  get fainter in both optical and X-rays, preventing the identification of higher redshift TDEs. Nevertheless, the bulk of the sample expected to be at $z \lesssim 0.5$ should still be successfully followed up at higher frequencies.
On the basis of these considerations, TDEs can be commensally searched for by exploiting the SKA1 all-sky surveys ($31,000 ~\rm deg^2$ at 2~$\upmu$Jy~beam$^{-1}$ at 1.5-2 arcsec resolution) proposed in the framework
of both the Magnetism and Continuum science cases (see Johnston-Hollit et
al. 2015, this volume; Prandoni et al. 2015, this volume). By carrying out multiple visits of the fields, with a cadence of $\sim 10$ days over the two year period, it will be very efficient in finding extra-galactic transients down to $\sim$90~$\upmu$Jy~beam$^{-1}$ .

\subsection{Implications: rich science}

For each of the two chosen example classes of objects we see, therefore, that the rates for detection for SKA1-MID are rather high, indicating a very high scientific yield from radio transients searches, simply for known objects, {\em as long as we commensally search all data sets (see below)}. Of course we can also estimate rates for many other types of object, such as X-ray binaries (Corbel et al. 2015), GRBs (Burlon et al. 2015) or supernovae (Perez-Torres et al. 2015, Wang et al. 2015), but the above are already sufficient to demonstrate the high scientific return likely from the SKA in this area.

For both TDEs and FRBs the predicted rates are between 1 and 1000 per week for SKA1-Mid. As noted above, this will be on top of a sea of other, less dramatic, but scientifically interesting variability. A good comparison area to TDEs and FRBs, and indeed one which has attracted similar researchers, is GRB afterglows: transient events which were for a long time mysterious in their origin (and to some extent still are), are associated with the most extreme astrophysical conditions, and can be seen to cosmologically significant distances. GRB afterglows have generated a very large number of high profile papers and citations, and continue to do so $\sim 15$ years after their discovery, have motivated major space missions (e.g. {\em Swift}) and are a major area of interest for the LSST. Furthermore, multiple high-profile papers are still being written about {\em individual} events. It is therefore entirely reasonable to surmise that many major papers per year could arise from radio transient studies, and be one of the main sources of early SKA publications (certainly while the deepest surveys are still being built up). 

For SKA1-Low the rates are less certain. Large numbers of FRBs {\em might} be detectable, but this depends strongly on the scattering and also on the spectra of the objects (rates calculated in the 1 GHz band for SKA1-Mid do not suffer from these uncertainties). LOFAR is probably the best facility against which to compare rates, although LW(D)A (Lazio et al. 2010) and MWA (Bell et al. 2014) rates are also relevant. Variability searches with LOFAR across $\sim 1500$ deg$^2$ as part of the LOFAR Transients Key Science Project have shown that the sky is essentially static at flux densities above a few mJy (at 140 MHz) on timescales of weeks -- months. However, there is good evidence for one or more bright ($\geq$ Jy) $\sim 10$-minute transients in a high-cadence shallow monitoring of the north celestial pole at 60 MHz (Stewart et al. 2014). If SKA1-Low can really achieve a continuum sensitivity of a few $\upmu$Jy in an hour across 27 deg$^2$ (although source confusion may be a major issue here), and the LOFAR transient reported by Stewart et al. is part of a population which follows $N(>F) \propto F^{-3/2}$ relation (as expected for an isotropic homogeneous distribution of sources throughout a flat space), then we would expect to find 1000s of such events per day. We note that completely independent of the SKA Transients SWG (and therefore of the authors of this paper), Metzger, Williams \& Berger (2105) have predicted that SKA surveys at 0.1--1 GHz should detect thousands of radio transients associated with GRBs, TDEs and neutron star mergers associated with magnetar remnants.

\subsection{Existing rates and limits from blind searches}

Complementary to the rates estimated for specific classes of object such as those given for the FRBs and TDEs given above, we can also consider the rates and limits set by existing blind searches for radio transients. In the following we attempt to provide a comprehensive summary of the published rates and limits. We separate these into slow/image-plane and fast/beamformed, although -- as noted earlier -- the boundary between the regimes is rather blurred.

\subsubsection{Transient rates and limits from imaging}

\begin{figure}
\includegraphics[width=.99\textwidth, angle=0]{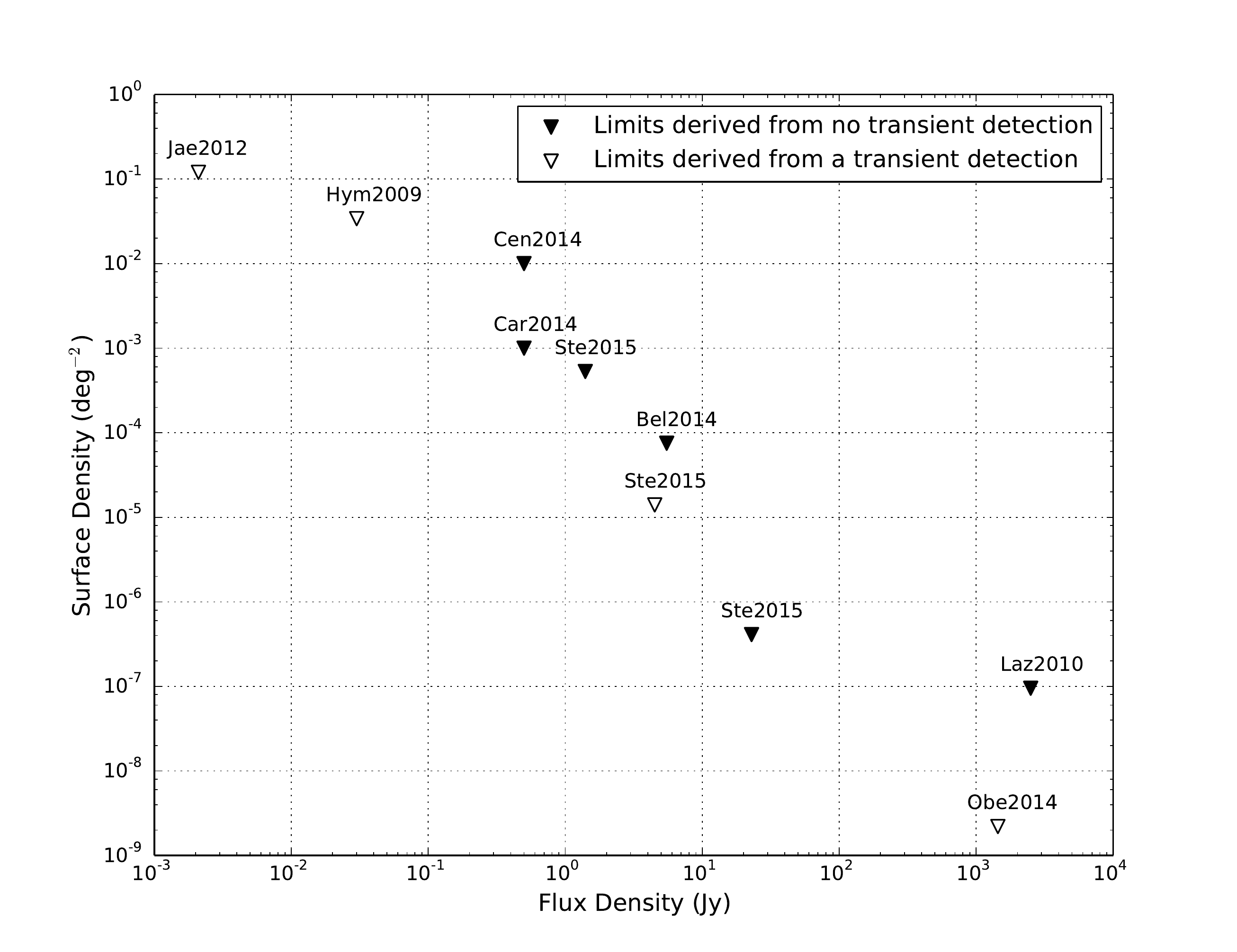}
\caption{A compilation of rates and upper limits for blind searches for radio transients at MHz frequencies. The surveys included in this plot are: Bell et al. (2014) (Bel2014); Carbone et al. (2014) (Car2014); Cendes et al. (2014) (Cen2014); Hyman et al. (2005,2006,2009) (Hym2009); Jaeger et al. (2012) (Jae2012); Lazio et al. (2010) (Laz2010), Obenberger et al. (2014) (Obe2014); and Stewart et al. (in prep) (Ste2015).}
\label{adammhz}
\end{figure}

\begin{figure}
\includegraphics[width=.99\textwidth, angle=0]{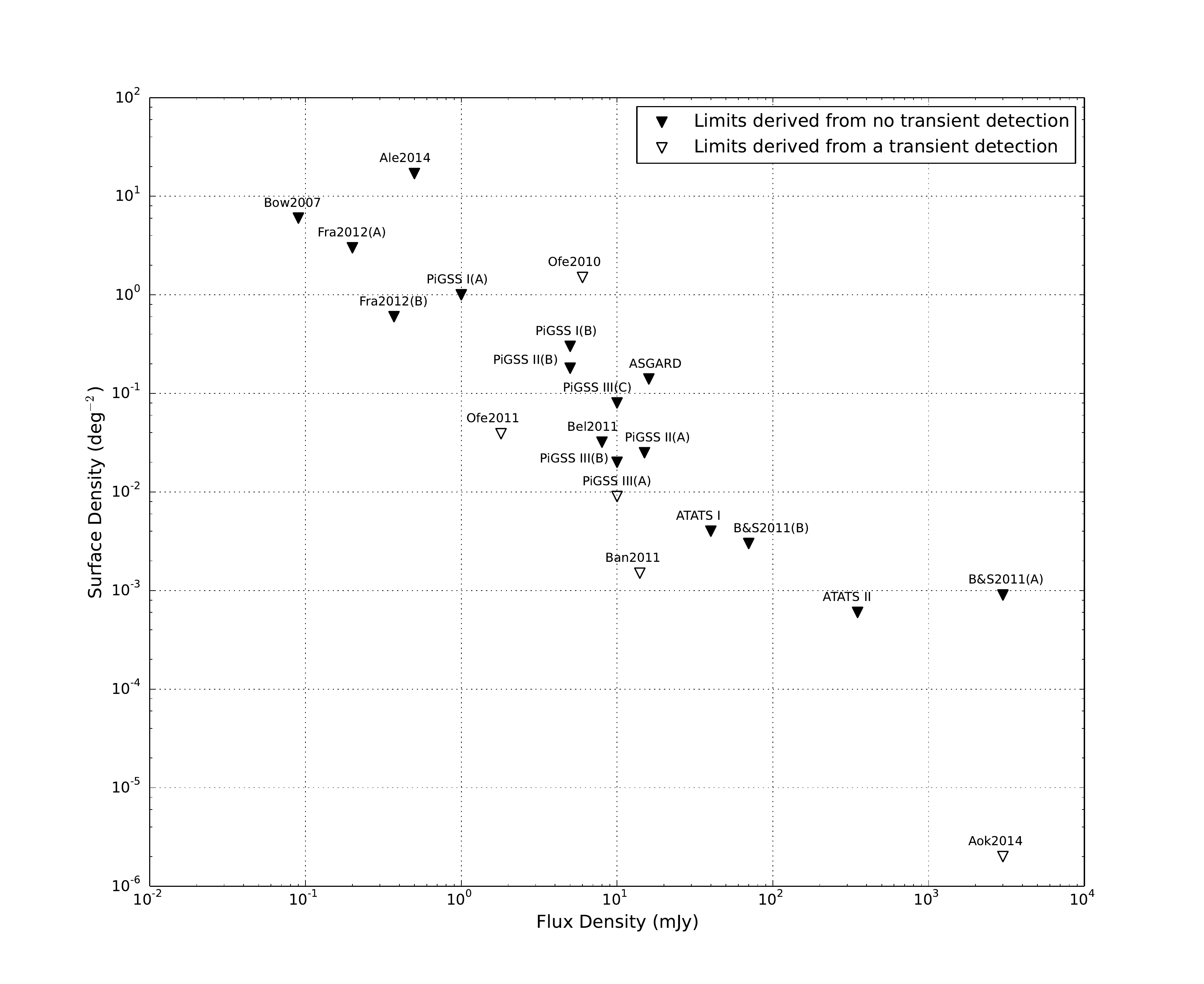}
\caption{A compilation of rates and upper limits for blind searches for radio transients at GHz frequencies. Multiple rates derived from the same survey are denoted by (A), (B) and (C). The surveys included in this plot are: Alexander et al. (2014) (Ale2014); Aoki et al. (2014) (Aok2014); Bannister et al. (2011a,b) (Ban2011); Bell et al. (2011) (Bel2011); Bower et al. (2007) (Bow2007); Bower et al. (2010) (PiGSS I); Bower et al. (2011) (PiGSS II); Bower \& Saul (2011) (B\&S2011); Croft et al. (2010) (ATATS I); Croft et al. (2011) (ATATS II); Croft et al. (2013) (PiGSS III); Frail et al. (2012) (Fra2012); Ofek et al. (2010) (Ofe2010); Ofek et al. (2011) (Ofe2011); and Williams et al. (2013) (ASGARD).}
\label{adamghz}
\end{figure}

A large number of image-plane surveys have been performed at a range of frequencies, observation timescales and cadences. Such rates are typically quoted as a surface density (limit), and these can be compiled and plotted as a function of flux density (corresponding to the sensitivity of the survey). A compilation of such limits, from image-plane surveys, presented in this way, is given in Table \ref{table:rates}, and Figs \ref{adammhz}, \ref{adamghz}. We caution that presenting the limits and rates in 2D plots such as these can be very misleading: observation timescales and cadence are also very important parameters for such surveys. We have tried presenting the data from Table \ref{table:rates} in a 3D form (such as that envisaged by Fender \& Bell 2011) but it is our feeling that the resulting plots confuse more than they clarify. Clearly in the future multi-dimensional analysis and Monte Carlo simulations will be required to determine how well parameter space has been explored.

\begin{table}
\scriptsize
\centering
\begin{tabular}{|c|c|c|c|c|c|c|}
\hline
\textbf{Survey/Paper}&Flux Limit& $\rho$&$\delta t$&$\Delta t$&$ \nu$&$N_e$\\
&(mJy)& (deg$^{-2}$) & & &(GHz)&\\
\hline
Stewart et al. (in prep)(A) & $ 2.25 \times 10^{4} $ &$4.1 \times 10^{-7}$ & 30 sec&cont - 4 months& 0.060 & 41\,350\\
Stewart et al. (in prep)(B)$^{*}$& $ 4\,100 $ &$1.4 \times 10^{-5}$ & 11 min&4 min - 4 months& 0.060 &1\,897\\
Stewart et al. (in prep)(C) & $ 1\,400 $ &$5.3 \times 10^{-4}$ & 297 min&4 min - 4 months& 0.060 &32\\
Lazio et al. (2010) & $ 2.5 \times 10^6$ &$9.5 \times 10^{-8}$ & 5 min&2 min - 5 min& 0.074 &$\sim$1\,272\\
Obenberger et al. (2014)$^{*}$ & $ 1.44 \times 10^6$ &$2.2 \times 10^{-9}$ & 5 sec & cont - year & 0.074 &$\sim$43\,056\\
Cendes et al. (2014) & $ 500$ &$10^{-2}$ & 11 min&min - months& 0.149&26\\
Carbone et al. (2014) & $ 500$ &$10^{-3}$ & 11 min&min - months&0.150&151\\
Bell et al. (2014) & $ 5\,500 $ &$7.5 \times 10^{-5}$ & 5 min&min - 1 year&0.154&51\\
Hyman et al. (2009)$^{*a}$ & $ 30 $ & 0.034 & $\sim$3 hr & days - months & 0.235, 0.330 & - \\
Jaeger et al. (2012)$^{*}$ & $ 2.1$ & 0.12 & 12 hr&1 day - 1 month&0.325&6\\
 & & & & & & \\
Bannister et al. (2011a,b)$^{*}$ &  14 & $1.5\times10^{-3}$ & 12 hr & 1 day - 20 yr & 0.843 & 2\\
Ofek et al. (2010)$^{*}$ & $$ 6 & 1.5 & days & - & 1.4 & 2\\
ATATS I/Croft et al. (2010) & $$ 40 & 0.004 & months & 1 month & 1.4 & 2\\
ATATS II/Croft et al. (2011) & $$ 350 & $6.0\times10^{-4}$ & ~ 1 day & min - days & 1.4 & 12\\
Bower \& Saul (2011)(A) & $$ 3000 & $9.0\times10^{-4}$ & 2 min & $>$ 1 min & 1.4 & 1\,852\\
Bower \& Saul (2011)(B) & $$ 70 & $3.0\times10^{-3}$& 2 min & 1 day - yrs  & 1.4 & 1\,852\\
Bell et al. (2011)& $$ 8 & 0.032 & 5 min & 4 - 45 days & 1.4, 4.9, 8.4 & 5037\\
Aoki et al. (2014)$^{*}$ & $$ 3000 & $2.0\times10^{-6}$ & 4 min & day & 1.4 & 1\,200\\
ASGARD/Williams et al. (2013) & $$ 16 & 0.14 & 1 hr & days - years & 3 & 29\\
PiGSS I/Bower et al. (2010)(A)& $$ 1 & 1 & months & months & 3.1 & 2\\
PiGSS I/Bower et al. (2010)(B) & $$ 5 & 0.3 & months & mins & 3.1 & 2\\
PiGSS II/Bower et al. (2011)(A) & $$ 15 & 0.025 & hours & days & 3.1 & 78\\
PiGSS II/Bower et al. (2011)(B) & $$ 5 & 0.18 & month & month & 3.1 & 5\\
PiGSS III/Croft et al. (2013)(A) & $$ 10 & 0.009 & hours & days & 3.1 & 379\\
PiGSS III/Croft et al. (2013)(B) & $$ 10 & 0.02 & months & months & 3.1 & -\\
PiGSS III/Croft et al. (2013)(C) & $$ 10 & 0.08 & months & years & 3.1 & 2\\
Bower et al. (2007) & $$ 0.9 & 6 & 20 min & years & 4.8, 8.4 & 17\\
Frail et al. (2012)(A) & $$ 0.2 & 3 & 20 min & month & 4.8, 8.4 & -\\
Frail et al. (2012)(B) & $$ 0.37 & 0.6 & 20 min & mins - days & 4.8, 8.4 & -\\
Ofek et al. (2011) & $$ 1.8 & 0.039 & 50 seconds & day - month & 4.9 & 16\\
Alexander et al. (2014) & $$ 0.5 & 17 & 15 mins & days - months & 4.9 & 31\\
\hline
SKA-Shallow & $$ $1.0 \times 10^{-3}$& $3.0 \times 10^{-8}$ & 4 hr & - & 1.4 & $\sim$5000\\ 
SKA-Deep &$$ $2.0 \times 10^{-5}$& $6.0 \times 10^{-3}$ & 5 hr & - & 1.4 & $\sim$500\\
\hline
\end{tabular}
\begin{flushleft}
$^{*}$ Limit derived from a transient detection.\\
$^{a}$ Using approximate rate derived in Williams et al. (2013) which is inclusive of Hyman et al. (2005,2006,2009).
\end{flushleft}
\caption{Summary of image-plane radio transient surface densities, $\rho$, as found in the literature. Columns: Keeping with the format set by Ofek et al. (2011), $\delta t$ is the timescale of each individual epoch, $\Delta t$ is the timescale(s) between epochs and $N_e$ is the total number of epochs. The first block in the main section of the table is considered the low-frequency portion of surface densities, which incorporates surveys conducted at frequencies $\leq$ 330 MHz. The second block are the densities defined at mainly GHz frequencies. The second section of the table provides estimates of transient surface densities that will be achievable with the SKA. An entry containing `-' signifies that we were not confident in identifying the respective value from the literature. All these values are those used in Figures \protect\ref{adammhz} and \protect\ref{adamghz}.}
\label{table:rates}
\end{table}

\subsubsection{Transient rates and limits from beamformed/pulsar modes}

\begin{table}
\scriptsize
\centering
\begin{tabular}{|c|c|c|c|c|c|c|c|c|c|c|c|}
\hline
Survey/ & Freq. & B/width & Int. time & Freq. res. & $\sigma_{1-{ms}}$ & Beamsize & Dwell time & $N_{\rm beams}$$^{A}$ & $N_{\rm FRBs}$ & Refs \\ 
Telescope  & (MHz) & (MHz) & ($\upmu$s)  & (MHz) & (Jy) & (sq. deg.) & (sec) & & &\\
\hline
LOTAAS$^{*}$ (coh.) / LOFAR      & 135  & 32  & 0.495 & 0.012 & 1.4    & 0.28 &3600 & 119133 & 0  & 1 \\
LOTAAS$^{*}$ (incoh.) / LOFAR      & 135  & 32  & 0.495 & 0.012 & 2.4    & 34   &3600 & 1953   & 0  & 1 \\
GBNCC$^{*}$ / GBT        & 350  & 100 & 0.082 & 0.024 & 0.065 & 0.41   & 120 & 30000  & 0  & 2  \\
PALFA / Arecibo    & 1440 & 100 & 0.064 & 0.39  & 0.011 & 0.0027 & 176 & 35067  & 1  & 3,4  \\
Parkes-MB (Gal.) / Parkes     & 1370 & 288 & 0.25  & 3     & 0.07  & 0.043  & 2100& 34710  & 1  & 5  \\
Parkes-MB (Int. lat.) / Parkes     & 1370 & 288 & 0.125 & 3     & 0.07  & 0.043  & 265 & 155142 & 1  & 6,7  \\
HTRU-N$^{*}$ / Effelsberg & 1400 & 300 & 0.054 & 0.59  & 0.036 & 0.017  & 90$^{B}$ & 1260000 & 0 & 8  \\
HTRU-S                  / Parkes     & 1350 & 340 & 0.064 & 0.39  & 0.064 & 0.043  & 270$^{C}$ & 554333 & 5 & 9  \\
\hline
SKA1-MID$^{D}$     & 1400 & 400 & 0.064 & 0.2   & 0.007 & 0.00027 & 600 & 11000000 & -  & 10,11  \\
SKA1-LOW$^{E}$     & 300  & 100 & 0.064 & 0.024 & 0.02   & 0.004   & 600 & 5000000  & -  & 10,11 \\
\hline
\end{tabular}
\begin{flushleft}
$^{A}$Given by the number of beams per pointing, multiplied by the number of pointings.\\
$^{B}$For the all-sky section of the survey; longer dwells are used of 180 and 1500 seconds are used at intermediate and low Galactic latitudes respectively.\\
$^{C}$For the all-sky section of the survey; longer dwells are of 540 and 4200 seconds are used at intermediate and low Galactic latitudes respectively.\\
$^{D}$Assumes 2000 tied array beams (including stations within 900m diameter) and a survey of the Galactic plane covering a total of $\sim$3000 square degrees.\\
$^{E}$Assumes 500 tied array beams (including stations within 1200m diameter) and a total sky area of $\sim$20000 square degrees.\\
$^{*}$Survey is ongoing, observations and/or analysis are not completed.\\
REFS:
1: Unpublished, see Coenen et al. (2014) for pilot project, also {\bf www.astron.nl/lotaas/},
2:Stovall et al. (2014),
3:Deneva et al. (2009),
4:Spitler et al. (2014),
5:Manchester et al. (2001),
6:Edwards et al. (2001),
7:Jacoby et al. (2009),
8:Ng and the HTRU collaboration (2013),
9:Keith et al. (2010),
10:Keane \& Petroff (2015),
11:Macquart et al. (2015)\\
\end{flushleft}
\caption{Overview of beamformed searches for radio transients.  The $\sigma_{1\mathrm{ms}}$ column gives the nominal 1-$\sigma$ sensitivity of the survey to a transient of duration 1 millisecond with a dispersion measure at which intra-channel smearing is negligible.}
\label{table:fastrates}
\end{table}

Table~\ref{table:fastrates} lists a sample of the largest beam-formed searches which have been previously performed and/or are currently ongoing. We caution that sensitivity estimates in this table were derived on the basis of nominal telescope characteristics and average sky characteristics in some cases, and that choices made during processing can have a substantial impact on the completeness of searches for short-duration, dispersed events Keane et al. (2015). Given the wide range of parameters encapsulated in these surveys we do not attempt to present them graphically. However, as noted above, the detection rate of FRBs should be more than a thousand times higher with SKA1-MID than Parkes, which has discovered the majority of FRBs to date.

\section{Optimising the SKA for transients}

In a set of meetings and discussions between 2013 -- 2015, the radio transients community concluded that while the telescope should be a superb facility for detecting radio transients, the major drawback of the existing designs was the lack of a requirement for (i) automatic, near real-time searching of data streams for transients, (ii) very fast (robotic) response modes for transients found in other ways.

\subsection{Commensality}

It is important to define commensal transient searches, and how we would want these to be implemented in the SKA system. By `commensal', we mean having no effect on (i) the original scientific goals and (ii) proposed observing strategy of an accepted proposal. Clause (i) is very strong and should be considered inviolate (of course we also hope that observations can be overridden by T-o-O style triggers but that is dealt with in the next section). Clause (ii) should be considered a default position but could potentially be modified at some stage if it increased the secondary science without harming the primary (consider a scenario in which we have discovered that transients which vary on $\sim$day timescales are extremely interesting -- it would then make sense to schedule a deep 8-hr pointing of some field as two 4-hr observation separated by one day). The implementation of commensal searches should, in our view, be part of the default SKA system and is not associated with custom instrumentation or `spigots'{\footnote{spigots have a particular definition within the SKA system design and should be not used lightly}}.

\begin{figure}
\includegraphics[width=.99\textwidth, angle=0]{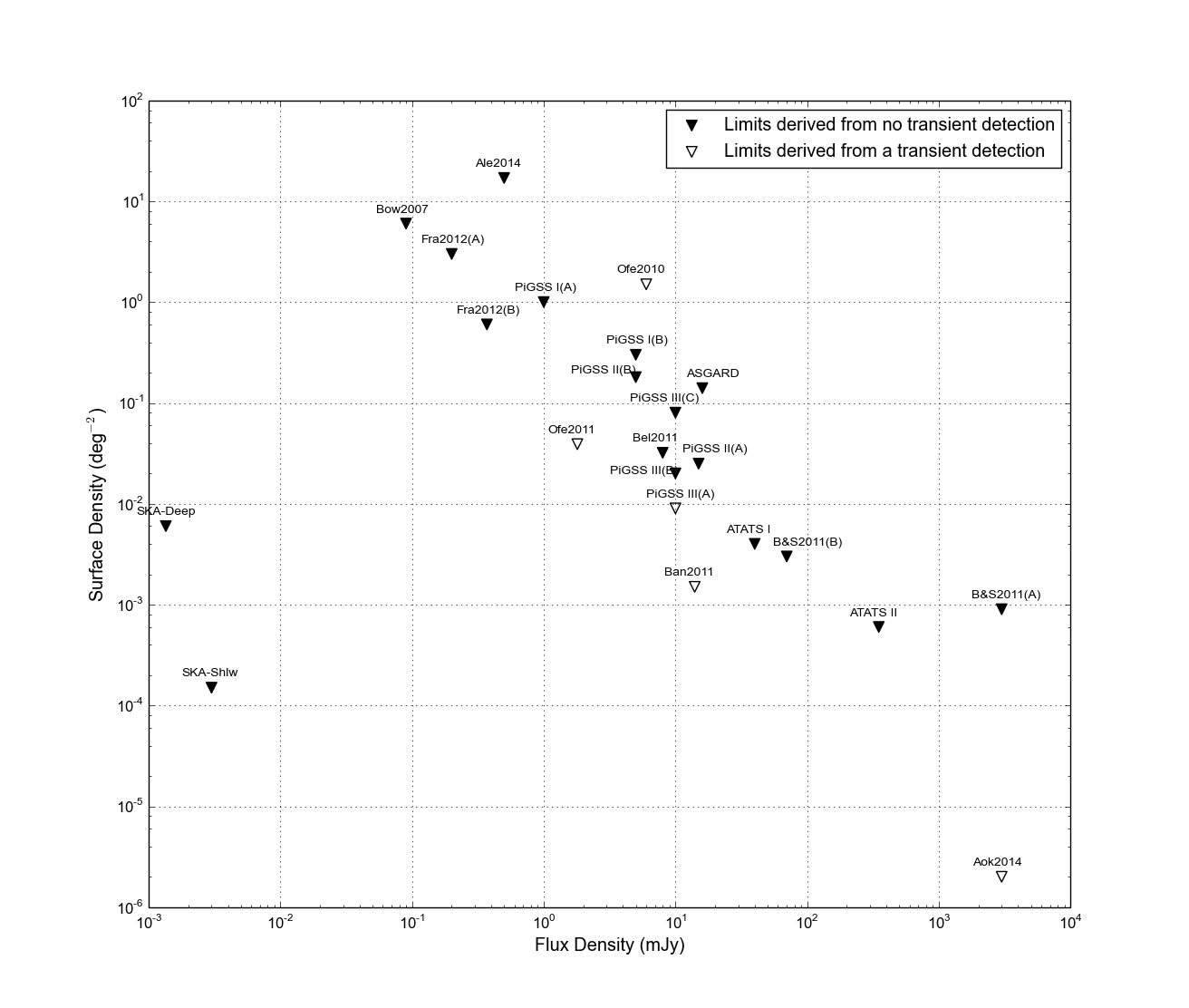}
\caption{Same as Fig \protect\ref{adamghz} with the survey sensitivities which would be achieved by searching commensally through the planned SKA-1 wide-and-shallow and narrow-and-deep continuum surveys. These limits correspond to one, and only one, way of searching these data: the same data can be searched on a variety of timescales in such a way theat they move more steeply than $-3/2$ in this representation. Specifically, for the estimated wide-and-shallow survey we consider two one-hour pointings, reaching $\sim 1 \upmu$Jy r.m.s., at each of 20 000 square degrees (one pointing per square degree). For narrow-and-deep we consider 500 pointings of $\sim 5$hr each on the same field, reaching an rms of about 450 nJy per pointing (ultimately giving $\sim 20$ nJy sensitivity across one square degree).}
\label{adamghzska}
\end{figure}


How can we compare the likely scientific harvest of `normal' (a.k.a. `conservative' or `20th century') observing with that of `fully commensal' (i.e. efficiently searching all data streams for transient events)? Considering briefly the `wedding cake' approach of most proposed/envisaged surveys (different tiers, with the shallowest being the widest), these are not very different from the search strategies a transients team would develop for the telescope if it had 100\% of the observing time. How much time is a transient programme likely to get in reality? We could imagine, optimistically, that this might be 10\%. Therefore, {\em if an efficient commensal transient search can be performed on all data streams}, then the rate of events increases by at least one order of magnitude. Fig \ref{adamghzska} is plotted in the form of Fig \ref{adamghz} but showing the rates/limits which would be achieved by piggybacking on the deep-and-narrow and wide-and-shallow continuum surveys which have been proposed for SKA1-MID (and would take place over the first few years of operation of SKA phase 1). These are vast improvements on current limits, which could be achieved 100\% commensally.
The FRB and TDE searches discussed above could also be achieved almost entirely commensally.
Facilitating the rapid commensal search of these data, as they are taken, for transients would not
only maximise the scientific yield on the telescope, it would also reduce the proposal time pressure.

Of course this is a big if, as is the associated design and implementation cost. Nevertheless, it is our judgement right now that such an addition to the telescope design would not incur a large fractional cost (and certainly much less than the rescope associated with the decisions taken in March 2015).

Two scenarios might be considered for commensal searching:

\begin{itemize}
\item{\bf Slow commensal:} searching of data for transients some time after the data have been taken, possibly when they have been moved from the current `live' data store to the archive. In many cases this timescale would be much longer than the timescale of the event and so, while easier to implement, this would have a lower scientific yield. Nevertheless, for many
luminous synchrotron events, timescales (for both latency and imaging) of $\sim$days, would not
seriously damage the programme. The computational demand of such a programme would
be entirely minimal.

\item{\bf Fast commensal:} Ideally, to fully explore parameter space and catch the fastest transients while counterparts at other wavelengths were still detectable, a near-real-time system could be implemented into the SKA {\em from the design phase} which was able to make a search for transients in data streams in close to real time (which here probably means timescales of
$\sim$seconds for latency and imaging). The computational demand for such a system would
be moderate, although the storage associated with continuously-generated 1sec, full-field
full-resution images would be significant and would mean that the transient search algorithms
would have to be working very fast to prevent a backlog.
\end{itemize}

\begin{figure}
\includegraphics[width=.99\textwidth, angle=0]{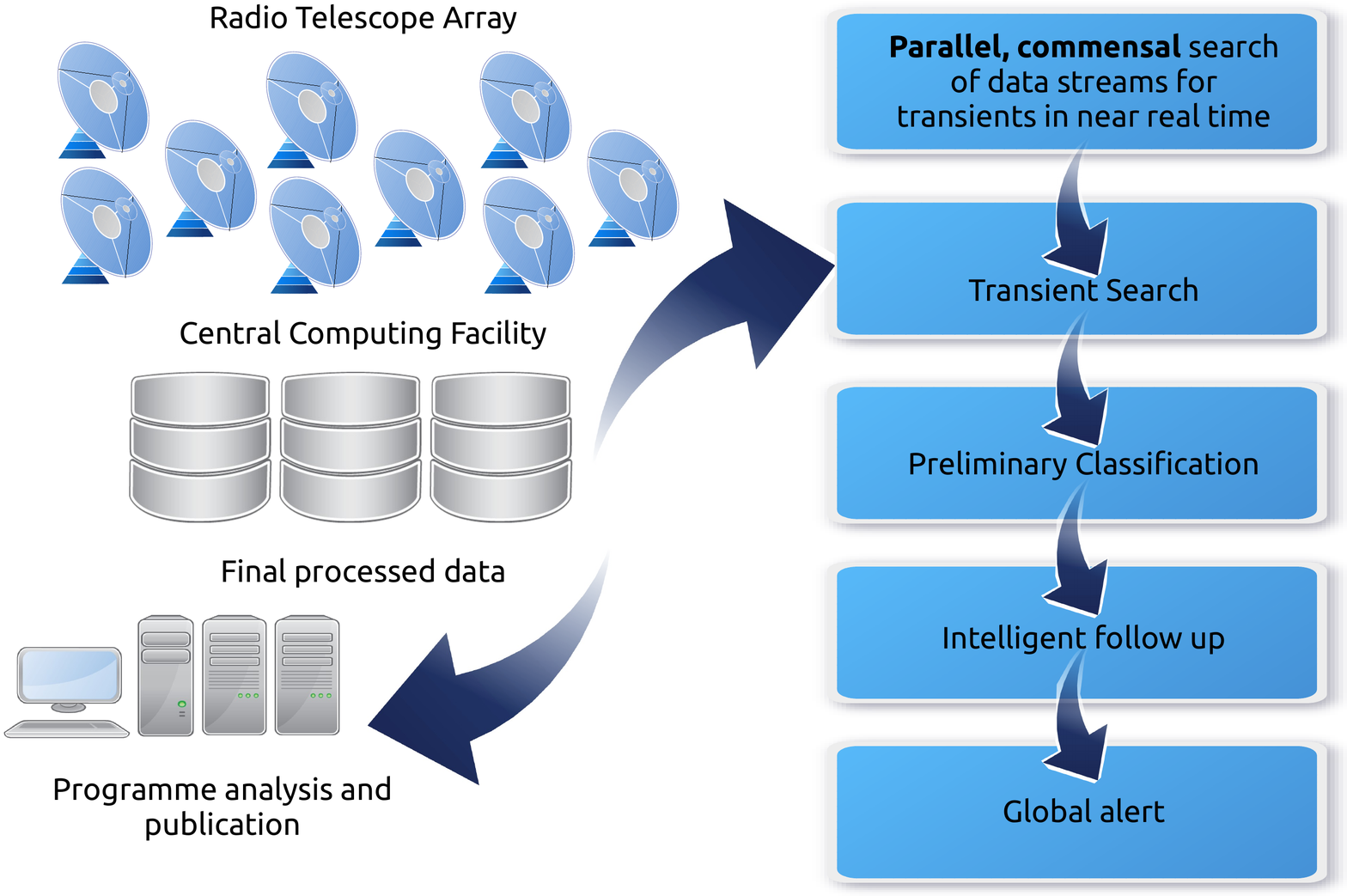}
\caption{Schematic of `traditional' and near-real-time transient searches occurring commensally, in parallel.}
\label{commensal}
\end{figure}

The conceptual parallel paths for near-real-time commensal transients searches and regular observing are illustrated in Fig \ref{commensal}. We remain convinced that it would be a missed opportunity not to include a fast-commensal system in the design of the SKA (certainly SKA1-Mid).

Of course the latter, fast commensal, system is the goal of our community, and we remain convinced that -- against the scale of the $\sim 30$\% project `rescope' in March 2015 -- it is entirely possible for it to be accommodated within a new baseline design. 

The politics of data ownership is also of course an issue here. Traditionally, but not exclusively, in astronomy, telescopes have awarded all the science in a data set to the proposers, for some proprietary period (typically 12 months). It is not obvious, however, that this is the right approach for facilities with very wide fields of view where multiple, distinct, science goals can be explored with the same data. In some such cases proposers are limited to certain science within a given data set -- such a system has been previously implemented for the gamma-ray observatory INTEGRAL. Specifically, for the first decade of INTEGRAL operations, a two-round system was used. In the first round, proposers could request certain fields to be observed, and define (and reserve) the primary science drivers for the proposal. In the second round, 'data rights proposals' could be used to obtain proprietary rights on other sources within the selected fields if they have not been reserved by the primary proposal.
This system has now been dropped, and a more traditional one-round system adopted, but this is primarily as a cost-cutting measure and the original system was widely considered to be successful with little evidence of gaming. However, a weakness of this system for dealing with transients was identified -- there was no guarantee that they would be spotted or acted upon in a timely manner (as with most telescopes), depending on the scientific inclinations and resources of the primary observers.

There is also a concern from some parts of the community that a small group would get to `own all the transients' and that this would be unfair. This is not the thrust of what we are proposing; rather we wish for the SKA to incorporate a generally-implemented and open system which delivers public alerts. Who then owns the data related to the follow-ups may be a question for a programme committee, but the global scientific community can only gain from immediate alerting to astrophysical transients. As noted above, this is precisely the policy adopted by several space missions, such as the enormously successful {\em Swift}, and is the planned policy for the LSST. How people deal with very large rates of transients, including how to select those which are potentially interesting and which to follow up with other facilities is another question. 

Specifically, therefore, we propose that the commensal system itself be transparent (i.e. publically available, up-to-date, documents and code) and that automatic alerts resulting from it are made globally and publicly, probably using the {\em VOEvent} framework. Of course there would be a commissioning/training phase in which alerts were only shared with trusted partners, but it should be possible to complete such a phase on a timescale of $\sim$year. Furthermore, we note that some groups (Armstrong et al. {\em in prep}) are already beginning to design a system for MeerKAT, which should be well-tested by the time SKA1-Mid is being deployed. For SKA1-Low, with its rather different family of transients (coherent vs synchrotron) and response modes, it would be desirable to be able to test such a mode on LOFAR (or maybe MWA), but this has not yet been implemented.

Such concepts, while not at present embedded in the SKA design, are not radical. Several projects within the radio domain have attempted to achieve similar goals, for example VFASTR (Wayth et al. 2011), European VLBI Network is (partially) implementing a robotic rapid-response system, the JVLA is implementing a real-time commensal fast transient detection system (Law et al. 2014), the Allen Telescope Array searched many of its data sets commensally for transients (e.g. ASGARD, Williams et al. 2013), the SUPERB programme at Parkes is making near-real-time searches for FRBs, NASA's Deep Space Network is searching for radio transients (O'Dea et al. 2014). The first steps towards a commensal image-plane transient search system for MeerKAT are also under design (Armstrong et al.{\em in prep}). 

\subsection{Rapid Response}

Rapid response to transient astrophysical events, so called Target-of-Opportunity (T-o-O) modes, is of great importance in astrophysics. As an example, the NASA {\em Swift} space mission was designed explicitly for very rapid slewing in response to GRB triggers; the most expensive array of optical telescopes in the world, ESO's Very Large Telescope, has had a Rapid Response Mode implemented. Such observations always cause inconvenience for the observing programme that they interrupt, but this has been shown time and again to be acceptably manageable by replacing lost time ASAP, something that is easy to do in an automatically queue-scheduled system.

Before 2012, no robotic override system, in which T-o-Os were implemented without human intervention, had ever been operated on a radio telescope (although the scientific utility of such modes had been demonstrated in the optical and X-ray bands). 
However, such a system has now been implemented on the AMI-LA telescope, an array of 
eight 12.8m radio dishes, operating in the 15 GHz band, and located in the UK. The array now routinely begins slewing to automatically-generated triggers based upon {\em Swift} alerts (transmitted using the {\em VOEvent} protocol) on timescales of less than a minute, taking data on-target within five minutes (Staley et al. 2013; part of the `4 PI SKY' project). This AMI-LA Rapid Response Mode, ALARRM, has been extremely successful both as a demonstrator and for doing real science, including possibly the earliest-ever detection of reverse shock radio emission from a GRB (Anderson et al. 2014; see Fig 2) and a prompt radio transient associated with a gamma-ray superflare from a nearby young binary system (Fender et al. 2015). An extension of the system, including {\em Fermi} alerts, has been occasionally implemented on the LOFAR-UK station at Chilbolton (Breton, Karastergiou {\em private communication}), and there are hopes that it could eventually be deployed for the entire LOFAR array. Similar in concept to the ALARRM programme, but using a single dish in pulsar mode, Bannister et al. (2012) also followed up several {\em Swift} GRBs without human intervention.

\begin{figure}
\includegraphics[width=.99\textwidth, angle=0]{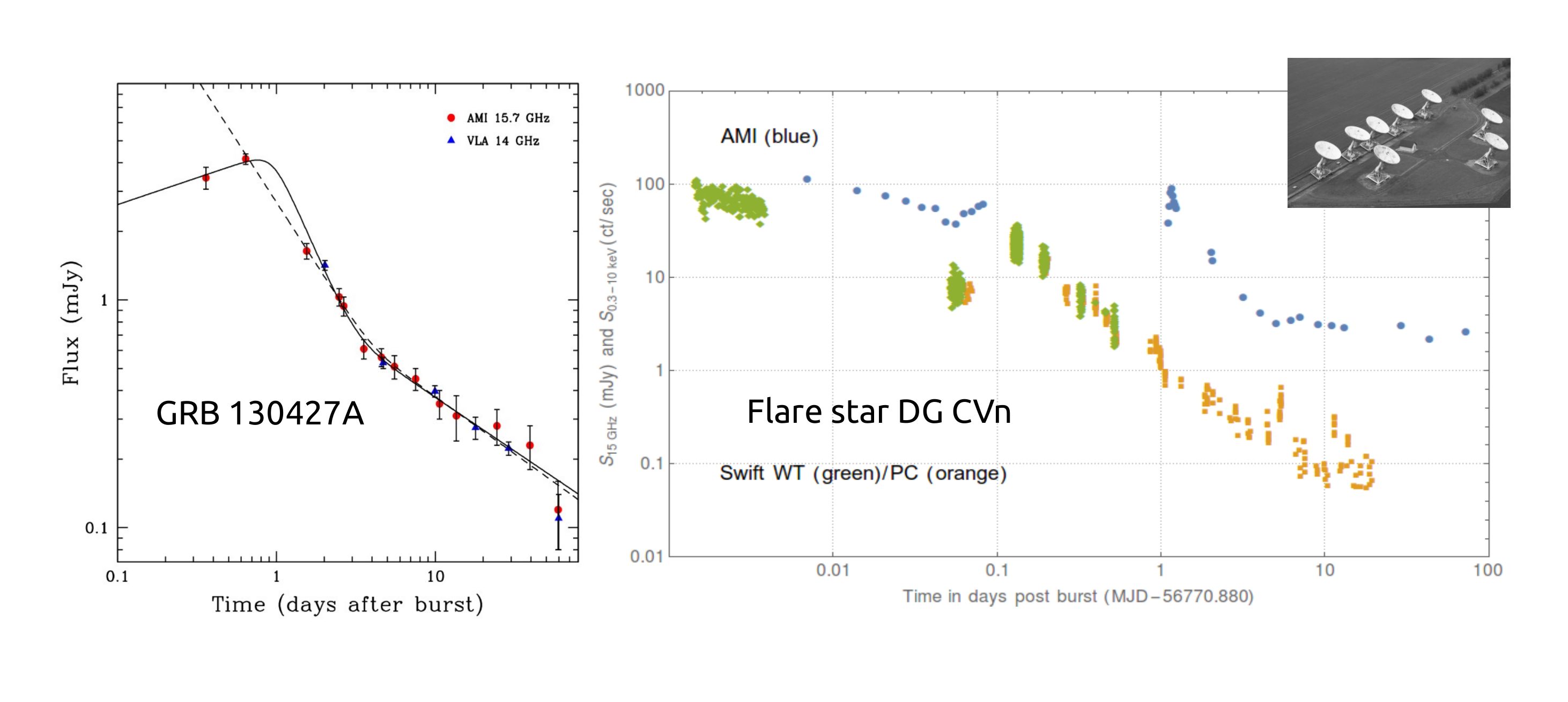}
\caption{Results from the ALARRM programme (see {\bf 4pisky.org}) in which the AMI-LA radio telescope has been configured to respond robotically (with no human interaction) to automatically-generated alerts from the {\em Swift} gamma-ray satellite.
{\bf Left panel:} AMI-LA observations of radio emission from Gamma Ray Burst GRB130427A, which at the moment of trigger was unobservable from AMI. Override observations were automatically scheduled and performed when the source was next at sufficient elevation, all without human intervention. The early time radio detections are amongst the earliest ever and probably detect the reverse shock in the early phases of the jet evolution (from Anderson et al. 2014).
{\bf Right panel:} Very early time radio flaring during a gamma-ray superflare from the young M dwarf binary DG CVn. In this case AMI-LA was observing within six minutes of the burst alert, and caught a $\sim 100$mJy flare which was already decaying (from Fender et al. 2015). The inset in the top-right hand corner is an aerial photo of AMI-LA.}
\label{ALARRM}
\end{figure}

In addition, the European VLBI Network (EVN) currently has two mechanisms for implementing rapid response observations. Besides the classical T-o-Os, the EVN has introduced a scheme of triggering proposals. In this case sources are observed if certain trigger condition (based on either an increase in the radio flux density, or changes e.g. in X-ray properties) are fulfilled. These projects can be triggered up to 1 day before real-time electronic VLBI (e-VLBI) observing sessions. There have been a number of automated trigger
tests in 2013/2014 to aim for much shorter timescale triggers, to be implemented in the near future. 

In summary, rapid response can be done, fast and efficiently, without human intervention, and we consider it essential that the SKA implement such modes.

\section{Summary}

The Square Kilometre Array has the potential to be a superb machine for the discovery of radio transients, yielding an extremely rich harvest of relativistic explosions, exotic astrophysics and previously unknown phenomena. As redesigned in 2015, the collecting area, wide bandwidth and spectral flexibility mean that the prospects for transients are extremely good. Nevertheless, a great opportunity would be missed, and the scientific harvest greatly reduced, if the following are not integrated into the system design:

\begin{itemize}
\item{{\bf Fast commensal searches for transients:} technically achievable, politically surmountable, this would increase the transient harvest by an order of magnitude, provide alerts for a global community hungry for interesting targets, and reduce the observer time pressure on the telescope.}
\item{{\bf Fast (robotic) response to transient alerts:} A further step forward can be made if the SKA can -- at times -- respond very rapidly to alerts generated either by other facilities or its own monitoring programmes (e.g. the commensal search, above). So much of the exciting astrophysics occurs early on in these phenomena that getting on source as early as possible (in reality, seconds to minutes) must be a high priority (for {\em some}, not all, events).}
\end{itemize}

We believe that the combination of the existing SKA design with these two features would make it a truly transformational facility for both stand-alone radio transients astrophysics and as part of a broad multiwavelength suite of telescopes exploring the most extreme phenomena the universe has to offer.

\section*{Acknowledgements}
Much of the knowledge in this work was distilled from countless meetings, telecons and email exchanges with the full and wide membership of the SKA Transients Science Working Group. We would like to specifically thank Tyler Bourke for assistance and patience in preparing this manuscript, and Evan Keane and Casey Law for useful comments and suggestions.

\end{document}